\documentclass{hotnets16}

\usepackage{times}  
\usepackage{epsfig}
\usepackage[TABBOTCAP]{subfigure}
\usepackage{tabularx}
\usepackage{graphicx} 
\usepackage{color}
\usepackage[dvipsnames]{xcolor}
\usepackage{xspace}
\usepackage{thumbpdf}
\usepackage{listings}
\usepackage{verbatim}
\usepackage{hyperref}
\usepackage{booktabs}
\usepackage{colortbl}
\usepackage{algorithm}
\usepackage{microtype}
\usepackage{booktabs}
\newcommand\subparagraph{%
  \@startsection{subparagraph}{5}
  {\parindent}
  {3.25ex \@plus 1ex \@minus .2ex}
  {-1em}
  {\normalfont\normalsize\bfseries}}
\makeatother
\usepackage[compact]{titlesec}
\let\subparagraph\relax

\usepackage[aboveskip=5pt, belowskip=-7pt, font={bf}]{caption}
\usepackage{multirow}
\usepackage{adjustbox}
\usepackage[T1]{fontenc}
\usepackage{adjustbox}

\usepackage{enumitem}
\setlist{itemsep=0.5pt,parsep=1pt,topsep=1pt,partopsep=1pt}

\hypersetup{pdfstartview=FitH,pdfpagelayout=SinglePage}

\begin{document}

\conferenceinfo{HotNets 2016} {}
\CopyrightYear{2016}
\crdata{X}
\date{}
\newcommand{\sys}{ExpressPass\xspace}

\newcommand{\ds}[1]{[\textcolor{blue}{\sf\textit{#1 - DH}}]}
\newcommand{\kj}[1]{[\textcolor{red}{\sf\textit{#1 - KJ}}]}
\newcommand{\inho}[1]{[\textcolor{cyan}{\sf\textit{#1 - Inho}}]}
\newcommand{\rev}[1]{\textcolor{black}{#1}}

\title{ExpressPass: End-to-End Credit-based\\ Congestion Control for Datacenters}

\author{Inho Cho$^*$, Dongsu Han$^*$, Keon Jang$^\dagger$ \\ \normalsize $^*$ KAIST, $^\dagger Google$}

\maketitle


\subsection*{Abstract}
As link speeds increase in datacenter networks, existing congestion control algorithms become less effective in providing fast convergence. TCP-based algorithms that probe for bandwidth take a long time to reach the fair-share and lead to long flow completion times. 
An ideal congestion control algorithms for datacenter must provide 1) zero data loss, 2) fast convergence, and 3) low buffer occupancy. However, these requirements present conflicting goals. For fast convergence, flows must ramp up quickly, but this risks packet losses and large queues. Thus, even the state-of-the-art algorithms, such as TIMELY and DCQCN, rely on link layer flow control (e.g., Priority-based Flow Control) to achieve zero loss. 
This paper presents a new approach, called \sys, an end-to-end credit-based congestion control algorithm for datacenters. \sys is inspired by credit-based flow control, but extends it to work end-to-end. The switches control the amount of credit packets by rate limiting and ensure data packets flow in the reverse direction without any loss. \sys leverages this to ramp up aggressively. \sys converges up to 80 times faster than DCTCP at 10Gbps link, and the gap increases as link speeds become faster. 
Our simulation with realistic workload shows that \sys significantly reduces the flow completion time especially for small and medium size flows compared to DCTCP, HULL, and DX. 

\section{Introduction}\label{sec:intro}
Datacenter networks are rapidly growing in terms of the size and link speed~\cite{singh2015jupiter}. A large datacenter network connects over 100 thousands machines using Clos network of small buffered switches~\cite{al2008scalable,greenberg2009vl2}. Each server is connected at 10~Gbps and 40~Gbps today with 100~Gbps on the horizon. This evolution enabled low latency and high bandwidth between servers within a datacenter. At the same time, this poses a unique and interesting  challenge for congestion control. 

In datacenters, short propagation delay makes queuing delay a dominant factor in end-to-end latency~\cite{alizadeh2010data}. With higher link speeds fast convergence has become much more important~\cite{jose2015high}. However, with buffers per port per Gbps actually getting smaller, high-speed links leave very little room for congestion control to adjust rates without incurring a packet loss and make fast convergence more difficult.
RDMA (Remote Direct Memory Access), recently deployed in datacenters~\cite{zhu2015congestion, azurerdma2016, mittal2015timely}, poses more stringent latency and performance requirements (e.g., zero data loss).

A large body of work addresses the above challenges. One such direction is to react to congestion early and more accurately either using ECN~\cite{alizadeh2010data, alizadeh2012less, zhu2015congestion} or delay~\cite{perry2014fastpass, mittal2015timely, lee2015accurate}. These approaches keep queuing lower and handle incast traffic much better than the traditional TCP\@. However, they are still prone to buffer overflows in bursty and incast traffic patterns.
Thus, they rely on PFC or avoid aggressive increase to prevent data loss. Another approach is proactive congestion control where the bandwidth of a flow or even the packet departure time is explicitly determined by a controller~\cite{perry2014fastpass, jose2015high}. However, it is very difficult to scale this approach to large datacenters and is challenging to make it robust.

We explore a new approach to ensure lossless and fast convergence while preserving the end-to-end principle without requiring changes to existing hardware. Our approach is inspired by credit based flow control~\cite{kung1994credit} used in Infiniband and high-speed system interconnect such as PCIe, Intel QuickPath, or AMD Hypertransport~\cite{Slogsnat:2007}. %
However, traditional credit based flow control is hop-by-hop, which requires switch support, and is difficult to scale to datacenter size.


Our approach takes the concept of credit and applies the end-to-end principle. We generate credit packets from the receiver to the sender on a per-flow basis, and let credit packets drop at the bottleneck on reverse path to determine the available bandwidth for data path.  By shaping the flow of credit packets in the network, the system proactively controls congestion even \emph{before} sending data packets. A sender gets to learn the amount of traffic that is safe to send by receiving ``credit'' packets, rather than reacting to the congestion signal generated from sending data packets. This ensures no congestion data packet losses, and allow us to quickly ramp up flows without worrying about data loss.




One might think 
that with the credit-based scheme a naive approach in which a receiver sends credit packets as fast as possible (i.e. the maximum credit rate corresponding to its line rate) can achieve fast convergence, high utilization, and fairness at the same time.
In the simplest case where $n$ flows share a single bottleneck, this is true.  
However, in large-scale networks, the three goals are often at odds, and the naive approach presents serious problems: ($i$) it wastes bandwidth when there are multiple bottlenecks, ($ii$) it does not guarantee fairness, and ($iii$)  latency difference between the time credit passed a bottleneck and the data packet arrival can create queuing and losses.
In addition, in networks with multiple paths credit and data packets may take asymmetric paths.



To address above challenges, we develop \sys that incorporates several techniques: ($i$) credit rate limiting at switches, ($ii$) symmetric hashing to achieve path symmetry, ($iii$) credit feedback control, ($iv$) random jitter, and ($v$) network calculus to determine maximum queuing. 

Our feedback control algorithm achieves fast convergence and zero data loss. It mitigates utilization and fairness issues in multi-bottleneck scenarios.
We quantify the tradeoff it achieves and show that the benefit often outweighs the cost.  Our evaluation using NS-2 simulation shows that it converges in just a few RTTs when a new flow starts for both 10 Gbps and 100 Gbps, whereas DCTCP takes over hundreds and thousands of RTTs respectively. In all of our simulations, \sys did not exhibit any single packet loss. \sys  used up to eight times less switch buffer than DCTCP, and data buffer is kept close to zero at all times.
Our simulation with realistic workload shows that \sys significantly reduces the flow completion time for small to medium size flows compared to DCTCP, HULL, and DX, and the gap increases with higher link speeds.

\section{\sys Design}\label{sec:design}

\begin{figure}[t]
\centering
\includegraphics[width=0.95\columnwidth]{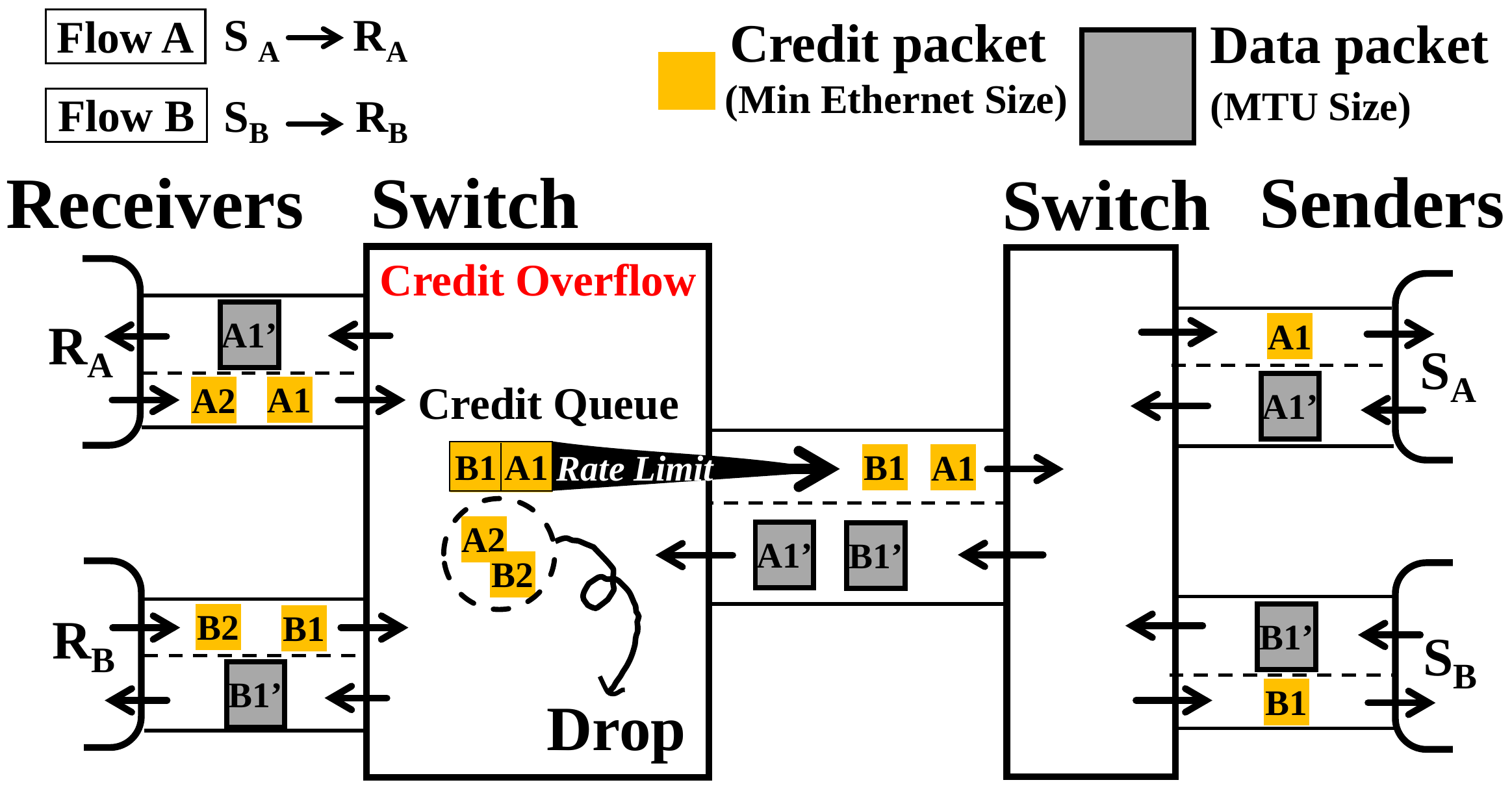}
\caption{\sys Overview}
\label{fig:c3-overview}
\end{figure}

This section illustrates how different components work together to make  credit-based congestion control work. 

\subsection{Basic approach}
\label{subsec:basic}
In \sys, credit packets are sent end-to-end on a per-flow basis. 
Each switch (and the host NIC) rate limits credit packets 
to ensure that the returning flow of data does not exceed the link capacity on a per-link basis. We assume symmetric routing that ensures data packets follow the reverse path of credit flows.
For credit packets, we use a 84 byte Ethernet\footnote{84B includes preamble and inter packet gap} frame and piggyback credit packets in TCP ACKs to minimize overhead. For each credit packet received, senders can send up to maximum Ethernet frame size of 1538 bytes including overhead. 
Thus, in Ethernet the credit is rate limited to 84/(84+1538) $\approx$ 5\% of the link capacity, and the remaining 95\%  is used for transmitting data packets. 
The credit throttling is applied per port, based on the link speed, thus having different link speeds in the network presents no problem as long as it is symmetric.

To illustrate how this mechanism allocates bandwidth, we draw a simple scenario with two flows in Figure~\ref{fig:c3-overview} where all links have the same capacity. Consider a time window in which only two packets can be transmitted on the link. Now, receiver $R_A$ and $R_B$ generate credits (A1, A2) and (B1, B2) respectively at the maximum credit rate. All credits arrives at the output port of the switch towards senders. Half of the credits need to be dropped at the output port as only two of them can go through due to throttling. In this example, A2 and B2 gets dropped, and each sender gets one credit. Each sender sends the data A1' and B1'. Note this generates exactly two data packets that can go through bottleneck link during the time window. This example demonstrates how throttling of credits allocates  bandwidth at a bottleneck link. 

\begin{figure}[t]
\centering
\includegraphics[width=1.0\columnwidth]{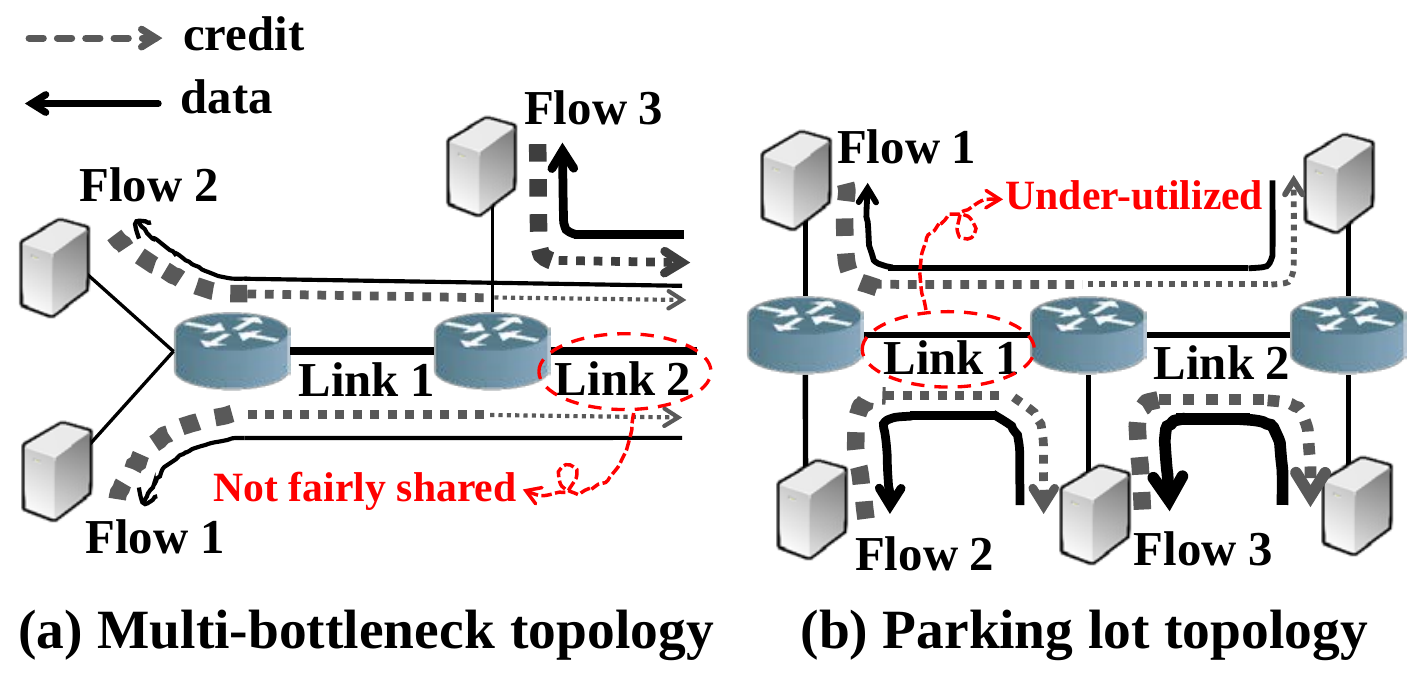}
\caption{Problems with naive credit based approach}
\label{fig:topology}
\end{figure}

However, naively sending credits at the maximum rate has two serious problems. First, it does not guarantee fairness. Consider the multi-bottleneck topology of Figure~\ref{fig:topology}-(a). When all flows send credit packets at the maximum rate, the second switch (from the left) will receive two times more credit packets for Flow 3 than Flow 1 and Flow 2. 
As a result flow 3 occupies two times more bandwidth on Link 2.
Second, multi-bottlnecks may result in low link utilization. Consider the  parking lot topology of Figure~\ref{fig:topology}-(b). When credits are sent at full speed, link 1's utilization drops to 83.3\%. This is because after 50\% of Flow1's credit passing link 1 (when competing with Flow 2), only 33.3\% of credit packets goes through Link 2, leaving the reverse path of Link 1 under-utilized. 

\vspace{0.05in}
\noindent
\textbf{Path symmetry:} Our mechanism  requires path symmetry---data packet must follow the reverse path of the corresponding credit packet. 
In datacenters with multiple paths (e.g., Clos networks), this can be done  by choosing the hash function to be symmetric for Equal Cost Multi Path (ECMP\@) routing. Two connected switches need to hash the credit and data packets of the same flow onto the same link (in different directions), where the same link may have different port IDs in the two switches. We omit the details. Note path symmetry does not affect performance of other schemes. Even with DCTCP, the utilization and performance on Fat-tree is not affected with path symmetry in our simulations.

\vspace{0.05in}
\noindent \textbf{Ensuring zero data loss:} 
Rate limiting credit packets controls the rate of data in the reverse path and make it practically congestion-free. However, different flows might have different RTTs. 
Thus, buffering is required to ensure zero data loss. To determine the exact amount of buffer to be lossless, we apply the network calculus~\cite{LeBoudec:2001} in the following fashion. Note, the time when credit came in through the port to a corresponding data packet arrives in the port is bounded by the network diameter and amount of queuing in the network. Suppose minimum and maximum delay between credit and data packet for a given port including queueing is $d_{min}$ and $d_{max}$ respectively. \rev{Both $d_{max}$ and $d_{min}$ are bounded by the network topology and amount of switch buffering.} The worst case arrival curve in terms of buffering is that for credit passed during $t = [0, d_{max} - d_{min}]$, all data packets arrives at the same time at $d_{max}$. This will create $d_{max} - d_{min}$ amount of buffering. We are working on the proof and evaluating whether existing switch provides large enough buffers to scale this to the size of datacenters. In our evaluation, with realistic workloads (Section~\ref{sec:evaluation}), the maximum switch buffer occupied by \sys is about one eighth of DCTCP\@.
When the buffer is even more scarce, \sys can operate with  smaller buffers by (i) lowering the credit queue length to minimize $d_{max}$ at the cost of under utilization, or (ii) relying on  PFC, like DCQCN and TIMELY, on the data path~\cite{zhu2015congestion,mittal2015timely}.

\vspace{0.05in}
\noindent\textbf{Starting and stopping credit flow:}
\rev{\sys requires a signaling mechanism to start the credit flow at the receiver. We piggyback credit request to either SYN or SYN+ACK packet depending on the data availability. This incurs  half an RTT  delay at the beginning. Persistent connections can send credit requests in a minimum size packet, but it adds overhead of an RTT\@. Credit request packet is sent as a regular data packet. Thus, it consumes data bandwidth without a credit and leads to data packet queueing. We are exploring options to either limit or eliminate such queueing. At the end of the flow, a sender marks the last data packet, and the receiver stops sending credits when it receives the last packet. This cause some credits to be wasted. We quantify the overhead and discuss mitigation strategies in Section~\ref{sec:discuss}.}

\subsection{Fast Convergence, Utilization and Fairness}
\label{subsec:fairness}
Fast convergence, utilization, and fairness present challenging trade-offs in congestion control. In our credit-based scheme, considering only one (e.g., fast convergence) results in an undesirable outcome as seen in the naive approach in Section~\ref{subsec:basic}. 
We describe the design of \sys that mitigates  undesirable outcomes (under-utilization and unfairness) while ensuring fast convergence.
However, when there is a hard conflict, we make a tradeoff in favor of the goals in the following order: fast convergence, high utilization, and fairness.


\begin{algorithm}[t]
\caption {Credit Feedback Control at Receiver.}
\label{alg:credit-feedback}
  \begin{itemize}
    \setlength\itemsep{0.01em}
    \item[] \textbf{(Initialization)} $cur\_rate = initial\_rate$
    \item[] Per update period (RTT by default)
    \item[] \quad \textbf{if} credit loss detected
    \item[] \qquad  $cur\_rate = average\_data\_rate$
    \item[] \quad \textbf{else}
    \item[] \qquad $cur\_rate = (cur\_rate+max\_rate)/2$
    \item[] \quad \textbf{endif} 
  \end{itemize}
\end{algorithm}

\vspace{0.05in}
\noindent
\textbf{Feedback control loop:}
\rev{On key component of \sys is feedback control. One question is: how does it differ from existing data packet feedback control? Feedback control on the data packets strives to balance between queueing, convergence, and utilization. By applying feedback control on credit packets, we remove queueing out of the equation. Because dropping credit packet is the norm, it simplifies the feedback control and enables a more aggressive adjustment.} 
\sys adjust the credit sending rate using the control loop described in Algorithm~\ref{alg:credit-feedback}.
It reduces the amount of credit if it detects any credit loss. Otherwise, it increase the credit rate. Detecting credit loss is straightforward. Each credit carries a sequence number, and data packet carries back the corresponding credit sequence number. If there is a gap in sequence numbers, we consider it a loss. Note, this is designed with the assumption of no packet reordering in the network. Removing this assumption is left as future work.

\sys adjusts rates aggressively for both increase and decrease to achieve fast convergence. 
During the increase phase, it converges to the midpoint between the maximum credit rate and current rate. This allows it to ramp up to over 90\% of link bandwidth in just 4 RTTs. In the decrease phase, we simply reduce it down to amount of credit delivered to the sender during the prior RTT, which is equal to the amount of data received by the receiver. \rev{One may think that a problem might arise when an application generates data at a lower rate than its fair-share bandwidth. However, our credit ``expires'' \emph{immediately} when no data is available at the sender. Thus, the receiver who sent the credit will detect credit ``loss'' and throttle the rate to applications demands.}

\begin{figure}[t]
\centering
\includegraphics[width=0.95\columnwidth]{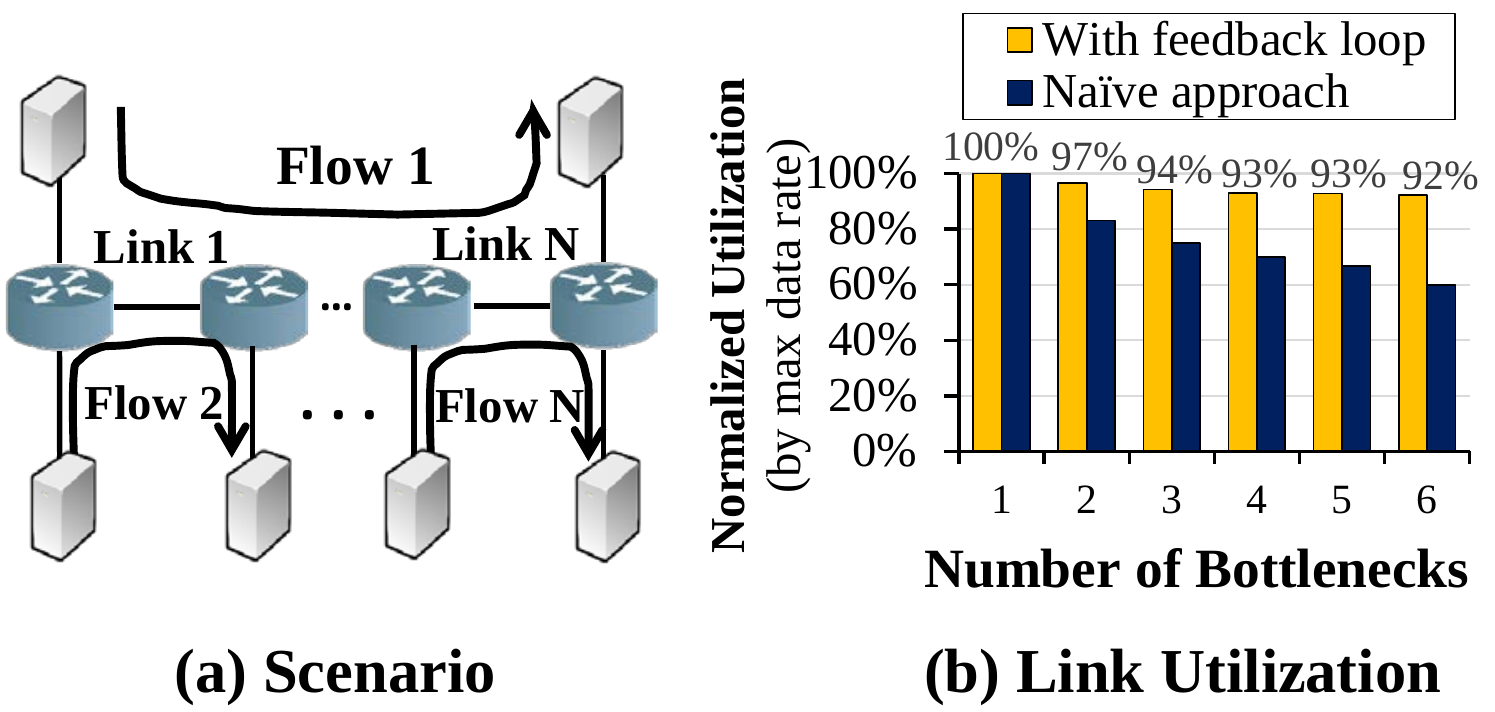}
\caption{Utilization in parking lot topology}
\vspace{-2pt}
\label{fig:utilization}
\end{figure}

\noindent
\textbf{Utilization with multiple bottlenecks:}
The feedback loop greatly improves utilization with multiple bottlenecks.
 Here, we quantify this using the topology of Figure~\ref{fig:utilization}. We increase the number of bottleneck links ($N$) from one to six.  Figure~\ref{fig:utilization} shows the utilization of the link that achieves the minimum utilization among Link 1 ... Link N. 
To isolate the loss due multiple bottlenecks, 
we report the utilization normalized to the maximum data rate excluding credit packets. As the number of bottleneck links increase, the utilization slowly drops. With two bottlenecks, the utilization improves to 96.7\% (from 83.3\% in the naive case) and with six bottlenecks, it achieves 92.2\% utilization (improvement from 60\%).


\begin{figure*}[t!]
\centering
    \begin{minipage}{0.38\textwidth}
        \centering
        \adjincludegraphics[width=\linewidth,trim={0 0 0 {0.13\height}},clip]{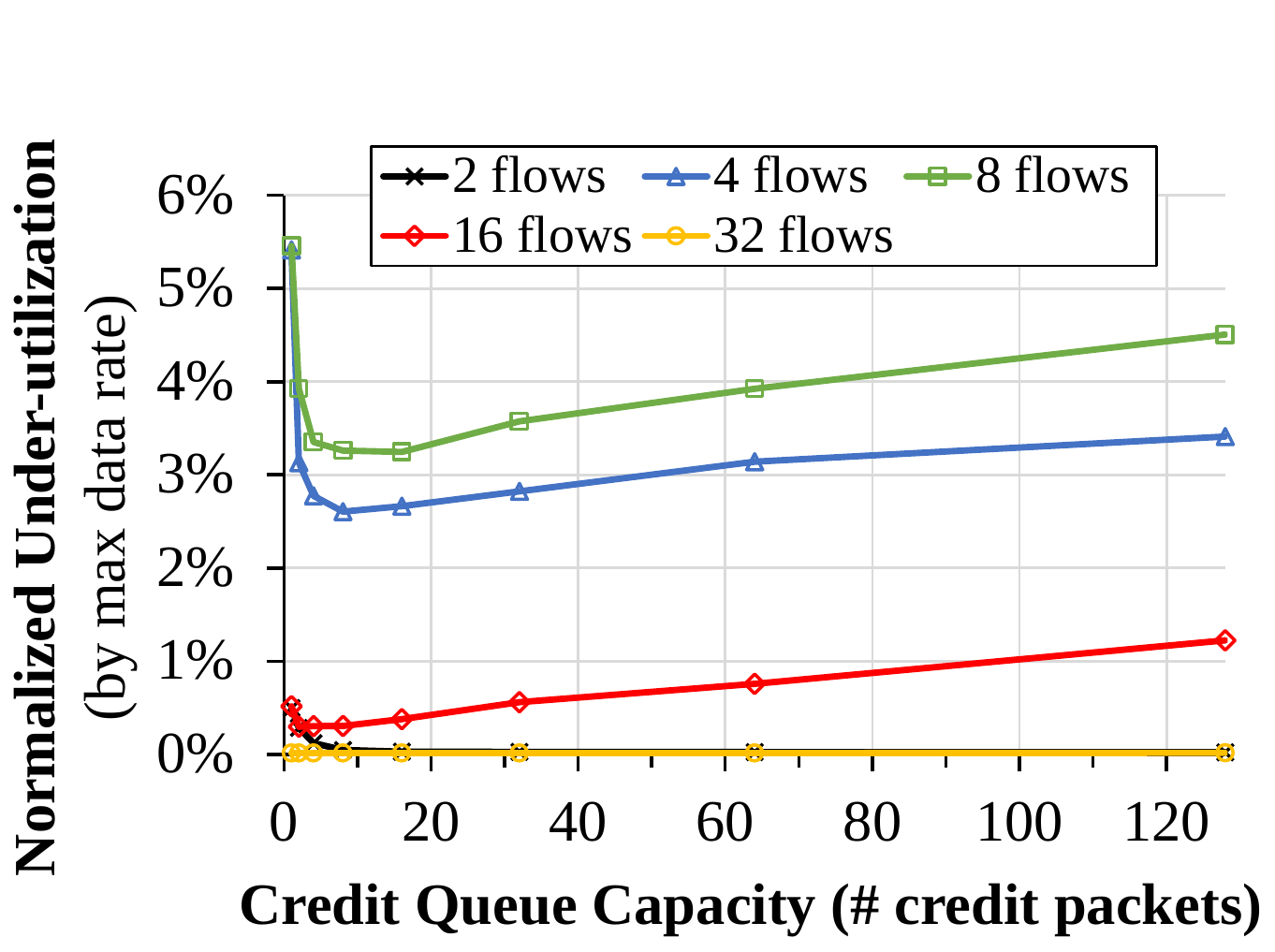}
        \caption{Credit queue size vs. Util.}
        \label{fig:c3_qlen}
    \end{minipage}
    \hspace{1.8cm}
    \begin{minipage}{0.38\textwidth}
        \centering
        \includegraphics[width=\linewidth]{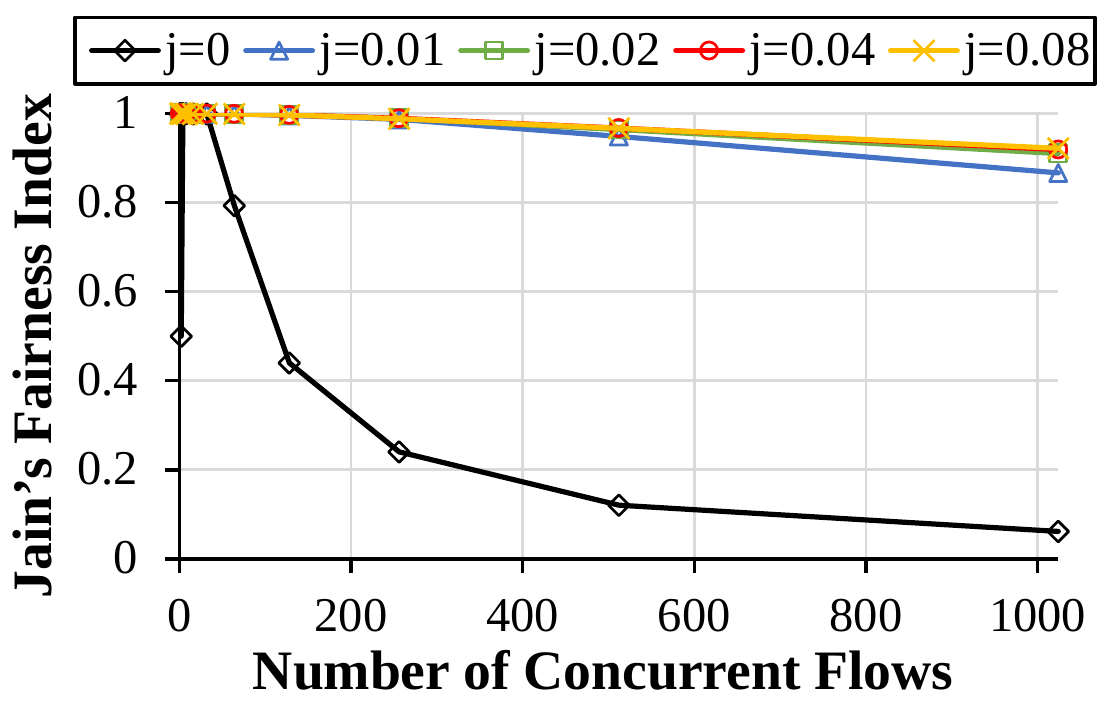}
        \caption{Jitter vs. Fairness}
        \label{fig:c3_jitter}
    \end{minipage}
    \vspace{2pt}
\end{figure*}

\begin{figure}[t]
   \centering 
        \includegraphics[width=\linewidth]{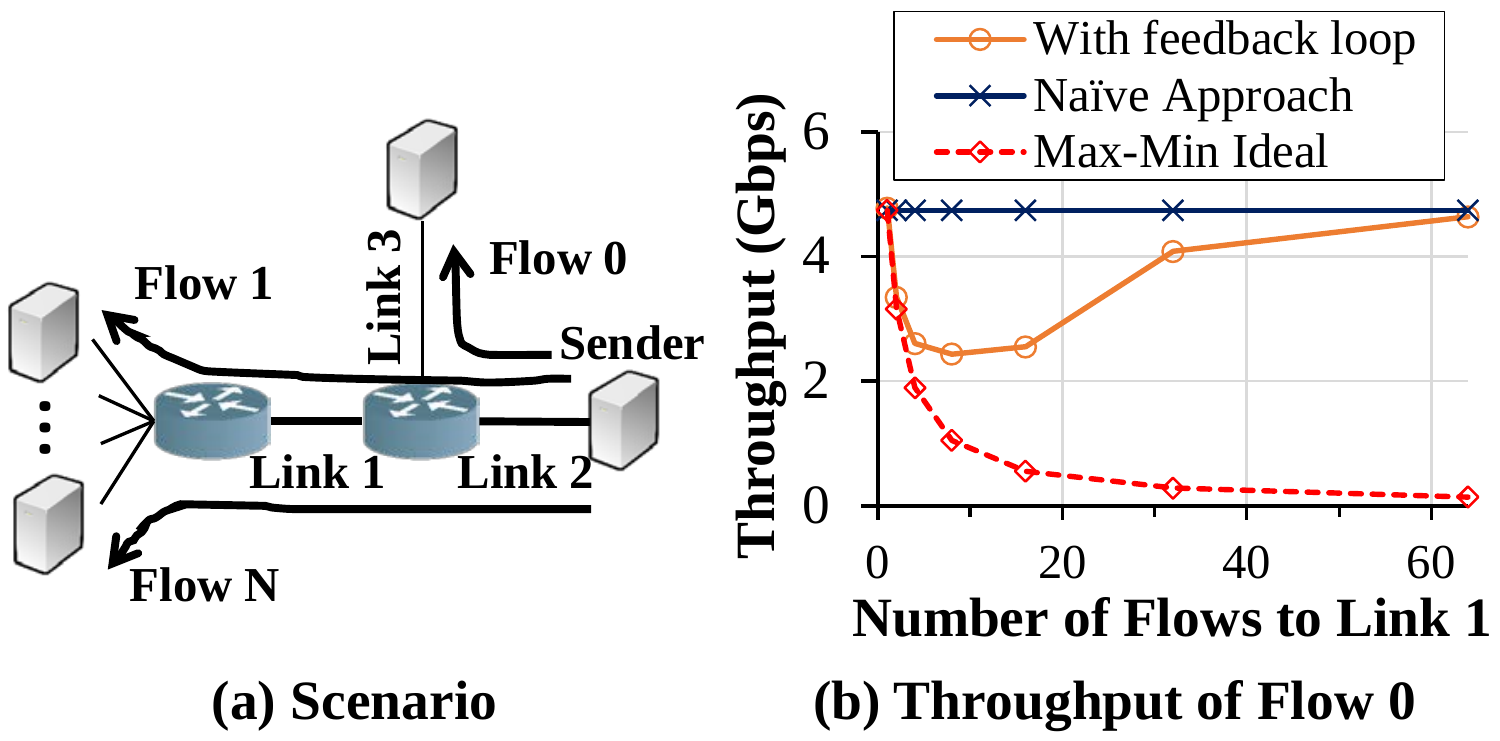}
        \vspace{-10pt}
        \caption{Fairness in multi-bottleneck}
        \label{fig:c3_mb_fairness}
        
\end{figure}

\noindent
\textbf{Credit queue size:}
\rev{Buffers hurt fast convergence as it delays feedback, but may be required for high utilization. In \sys data packets experience low buffer occupancy, because we throttle credit packets. However, now buffering may still occur with the credit packets. Thus, we quantify how much credit queue is necessary to ensure high utilization.} For this, we conduct a simple experiment using a dumbbell topology with our feedback control. 
We vary the credit queue size from 1 to 128 packets in power of 2 and measure the corresponding under-utilization.  Figure~\ref{fig:c3_qlen} shows that credit queue size affects utilization. The utilization maxes out at queue sizes between 8 and 16.  In the rest of the experiments, we use 16, which risks about 3.3\% of under-utilization in the worst case. 

\vspace{0.02in}
\noindent
\textbf{Convergence behavior:}
We give an intuition why the feedback control converge to fair-share in the single bottleneck case. Suppose there are two flows (A and B) with credit rate of $R_A$ and $R_B$ respectively. Assume ratio between them is $r = {{R_A} \over {R_B}} < 1$, meaning the convergence is not reached. During the decrease phase, with high probability $R_A$ and $R_B$ will be decreased by the same factor, not changing the ratio at all. Randomized jitter helps to achieve this property. However, during the increase phase, new ratio will be ${{(R_A + C) / 2} \over { (R_B + C) / 2}} = {{R_A + C} \over {R_B + C}}$ where C is $max\_rate$. Because, $ 1 > {{R_A + C} \over {R_B + C}} > {{R_A}\over{R_B}}$, the ratio converges towards $r=1$, and achieves fairness. In multi-bottleneck scenarios, our algorithm makes a trade-off, which we quantify this later in the Section.


\vspace{0.02in}
\noindent
\textbf{Randomized Jitter:} Our feedback control relies on uniform random dropping of credit packets---if $n$ flows are sending credits at the same rate to the shared bottleneck, equal fraction of credit packets must be dropped from each flow. Unfortunately, subtle timing issue can easily 
result in a skewed credit drop with drop-tail queues.
To address this issue, we rely on randomization. We introduce random jitter in sending credit packets, instead of perfectly pacing them. To evaluate the impact of jitter on fairness, we create a varying number of concurrent flows (1 to 1024) in a dumbbell topology. 
We vary the jitter level ($j$) from 0.01 to 0.08, relative to the inter-credit gap.  We then measure the fairness using Jain's fairness index~\cite{jain1984quantitative} over an interval of 1 msecs. Figure~\ref{fig:c3_jitter} shows the result, where fairness index of 1 means perfect fairness.
The result shows small jitter is enough to achieve good fairness.

\noindent
\textbf{Fairness trade-offs:}
Our feedback control improves utilization and fairness with multiple bottlenecks. However, it does not guarantee fairness at all times. To quantify the tradeoff, we use the multi-bottleneck scenario in Figure~\ref{fig:c3_mb_fairness} (a) and vary the number of flows that use Link 1. We then measure the throughput of flow 0. With ideal max-min fairness, flow 0 should gets 1/N of the link capacity (red dashed line in Figure~\ref{fig:c3_mb_fairness} (b)). \sys follows the max-min fairshare closely until four flows. But, as the number of flows increases, it converges towards the naive approach. This is because our rate control is too aggressive at increase. 
Finally, our feedback currently assumes host link speeds are the same (e.g., $max\_rate$ is the same across all flows), which may not be always the case. When host's link speeds differ, it also impacts fairness. 

We believe fairness and fast ramp-up are often at odds. On a much less aggressive side, we have
TCP's AIMD control. 
Applying AIMD to credit rate control would not require credit packets to drop proportionally. This may achieve fairness, but sacrifice fast convergence. This paper explores a much more aggressive end and demonstrates its viability. 
Exploring a feedback control loop 
that finds a better balance between fast convergence and fairness is left as future work.

\begin{figure}[t]
\centering
\includegraphics[width=1.0\columnwidth]{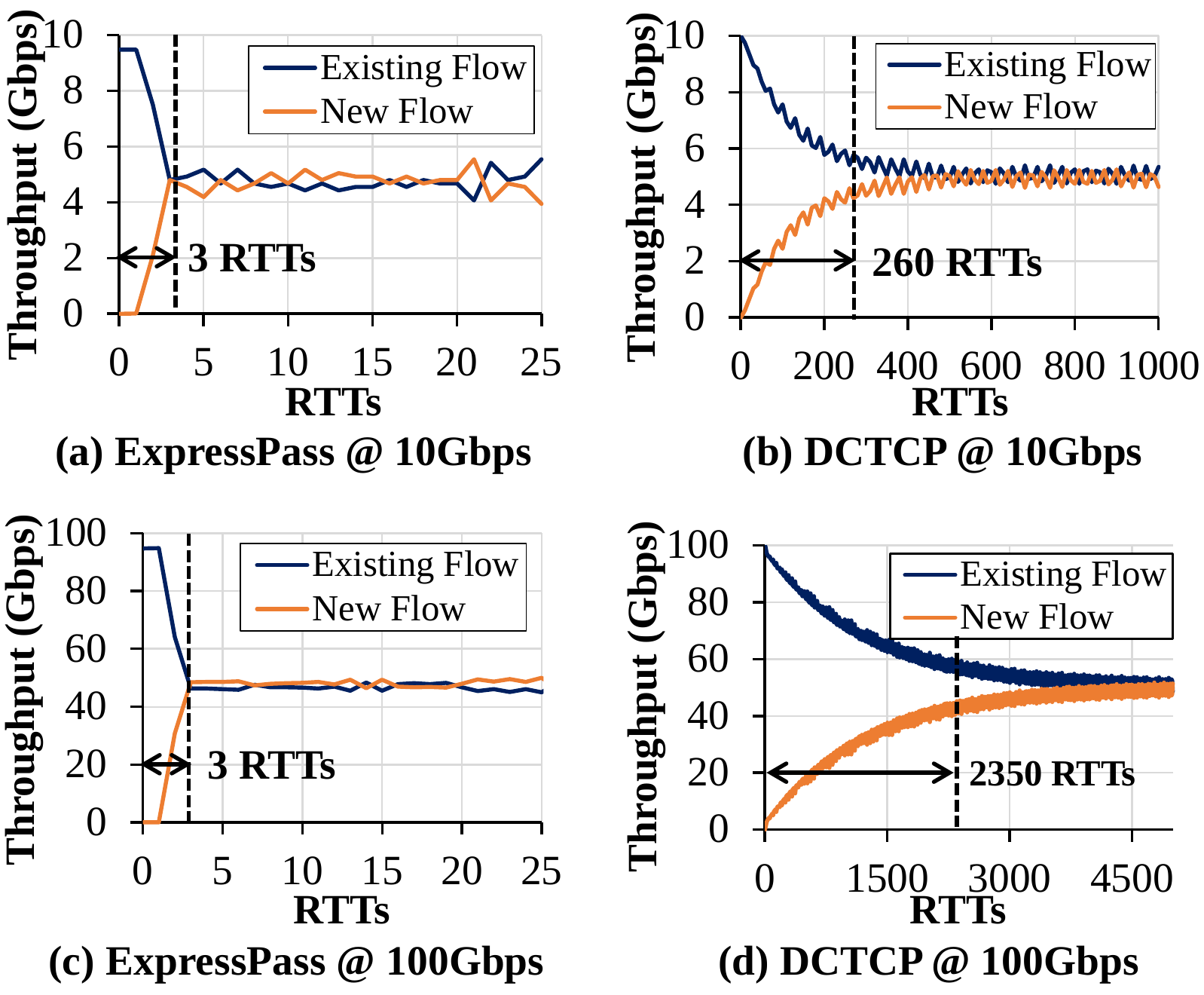}
\vspace{-12pt}
\caption{Convergence time with 100$\mu s$ RTT}
\label{fig:c3-conv}
\end{figure}

\begin{figure}[t]
\centering
\vspace{10pt}
\includegraphics[width=0.75\columnwidth]{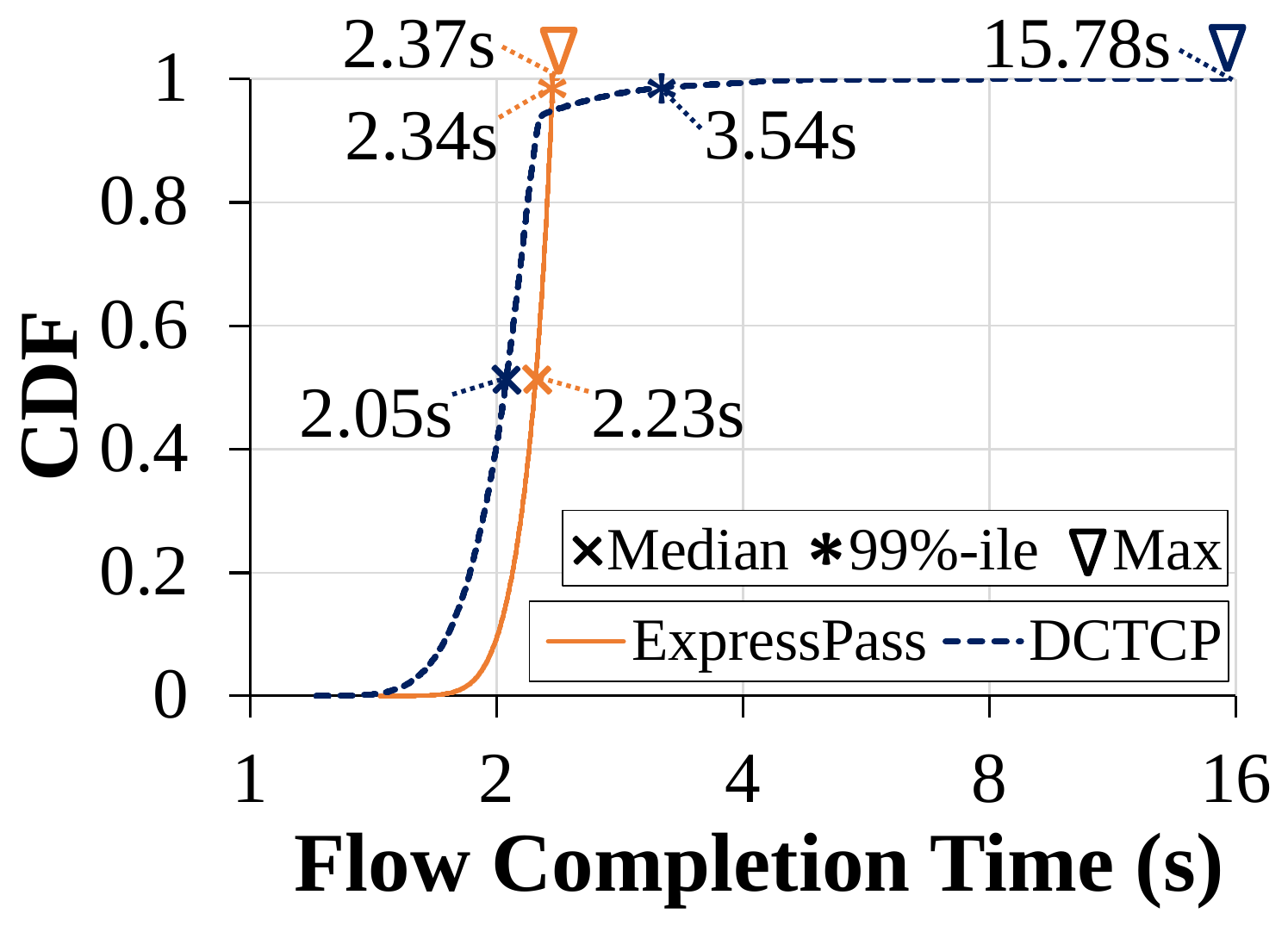}
\caption{Shuffle workload}
\label{fig:shuffle_fct}
\end{figure}


\section{Evaluation}\label{sec:evaluation}

We evaluate the following aspects of \sys using NS-2~\cite{mccanne1997network} simulation:

\begin{enumerate}
    \item Does \sys{} provide fast convergence?
    \item Does it provide high utilization and fairness?
    \item How does it perform under realistic workloads?
\end{enumerate}

For \sys, we use the credit queue size of 16 credit packets (1.34KB), and jitter of 1\% of inter-credit interval. We use  minimum RTO of 200 usec as suggested in ~\cite{vasudevan2009safe}. 

\subsection{Basic Performance Microbenchmark}\label{eval_convergence}

We start two flows one by one and measure how \sys and DCTCP\@ quickly ramp up to the fair-share. We define the convergence point as the time when two flows are within 10\% of the fair-share. We vary the bottleneck link speed from 10 Gbps to 100 Gbps. 
The RTT (without queueing) is set to 100 us. We set the DCTCP parameter K= 65, g= 0.0625 for 10 Gbps link, 
and K= 650, g= 0.01976 for 100 Gbps link.


Figure~\ref{fig:c3-conv} shows the flows' throughput for \sys and DCTCP at each RTT\@. \footnote{For \sys, we report the average throughput for each RTT. For DCTCP, we averaged over 10 RTT\@.}
\sys converges within 3 RTTs, while DCTCP takes more than 80 times longer than \sys with 10 Gbps link. As bottleneck link capacity increases, the convergence time gap between \sys and DCTCP becomes larger. At 100 Gbps, \sys's convergence time remains unchanged, while DCTCP's convergence time grows linearly to the bottleneck link capacity. Because of DCTCP's additive increase behavior, its convergence time is proportional to the bandwidth-delay product (BDP). Note that the maximum \sys data throughput is 94.82\% of the link capacity (9.482 Gbps on a 10 Gbps link) because 5.18\% of the bandwidth is reserved for credit.

\begin{figure}[t]
\centering
\includegraphics[width=0.98\columnwidth]{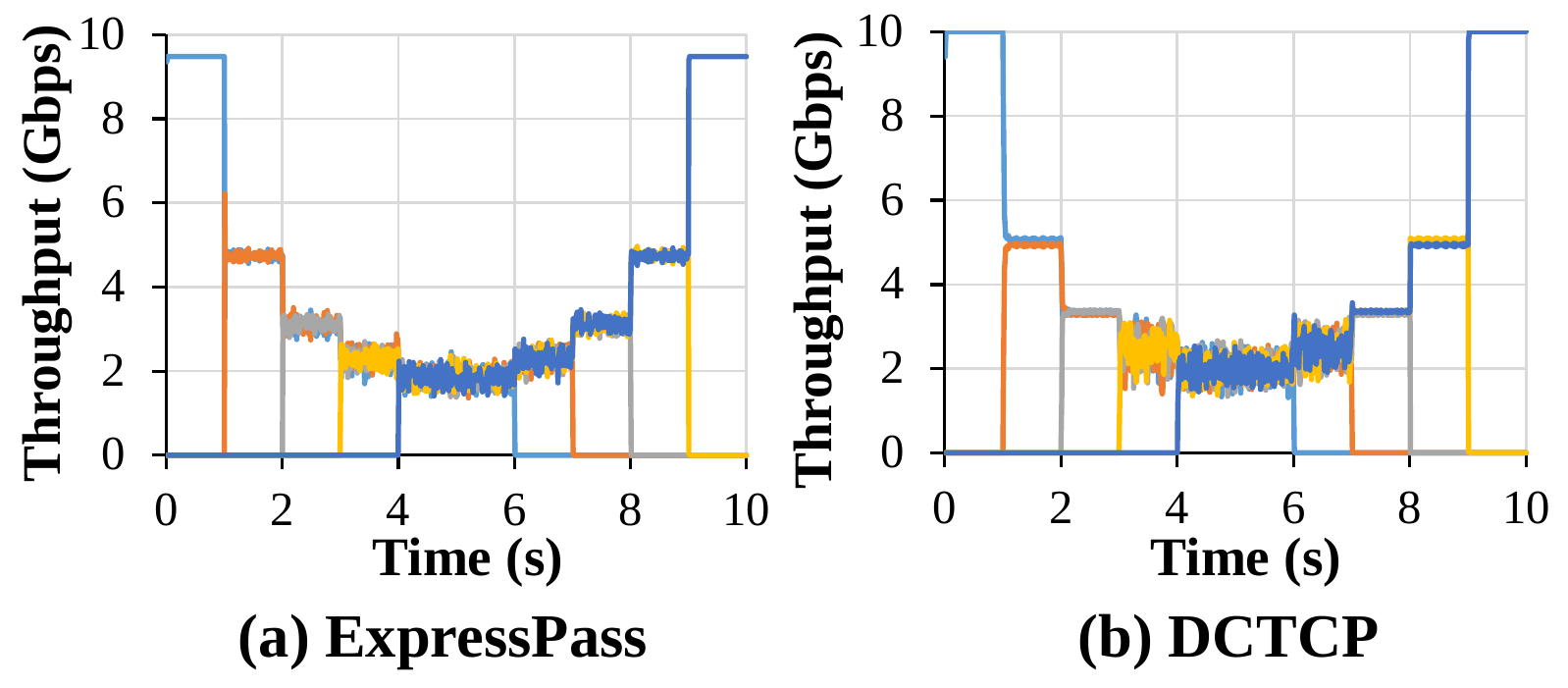}
\vspace{-0.1in}
\caption{Convergence behavior}
\label{fig:fairness}
\end{figure}

Figure~\ref{fig:fairness} shows the behavior with up to five flows joining and leaving over time for \sys and DCTCP\@. It shows that both \sys and DCTCP\@ can achieve fair-share.
\begin{figure*}[!t]
\centering
\includegraphics[width=\linewidth]{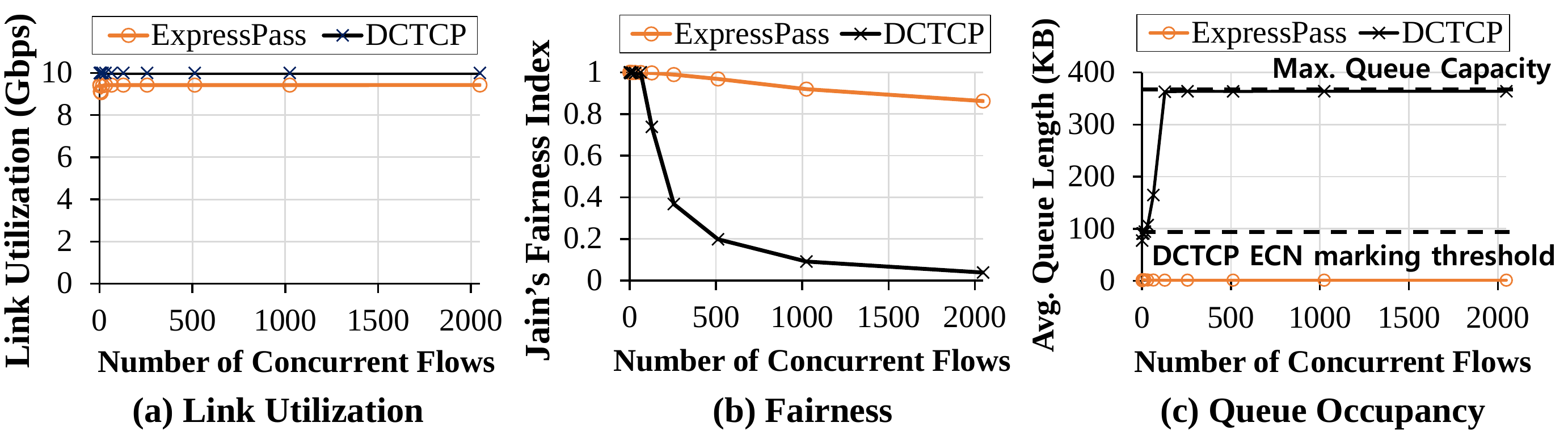}
\caption{Many concurrent flows in dumbbell topology}
\label{fig:manyflow}
\end{figure*}

\subsection{Heavy incast traffic pattern.}
One advantage of \sys is robustness against incast traffic pattern. Such traffic pattern happens common in shuffle step of MapReduce~\cite{dean2008mapreduce}. It creates an all-to-all traffic pattern, generating incast towards each host running a task. We model 40 hosts connected to single top-of-rack (ToR) switch via 10Gbps links. Each host runs 8 tasks, each of which sends $1$~MB to all other tasks. Thus, each host sends and receives $2496$ $(39 \times 8 \times 8)$ flows.  Figure~\ref{fig:shuffle_fct} shows the CDF of flow completion times (FCTs) with DCTCP and \sys. The median FCT of DCTCP is slightly better (2.0 vs. 2.2 sec). But DCTCP has a much longer tail. 
At 99-percentile and tail,  
\sys outperforms DCTCP by a factor of 1.51 and 6.65 respectively. 
With DCTCP, when some faster flows complete, the remaining flows often catch up. However, at the tail end, delayed flows tend to be toward a small set of hosts, such that they cannot simply catch up by using all available bandwidth. \rev{This drastically increases the tail latency
and contributes to the straggler problem in MapReduce~\cite{ananthanarayanan2010reining}.}
Our example demonstrates that network congestion control can contribute to the straggler problem and \sys can effectively alleviate this.



To better understand the underlying cause of the shuffle result, we characterize the behavior of both schemes with a large number of concurrent flows (up to 2,048) sharing a 10 Gbps bottleneck link. Figure~\ref{fig:manyflow} shows the utilization, fairness, and queue occupancy. We compute the Jain's fairness index using the throughput of each flow at every 100 msec interval and report the average. Both DCTCP and \sys provide high utilization. However, with larger number of flows DCTCP's fairness drops significantly (to 0.04 at 2,048 flows). This is because DCTCP cannot handle the congestion window of less than 1; some flows time out and collapse.  In contrast, \sys provides better fairness, shows near zero data queuing, and uses much less buffer than DCTCP.

\subsection{Large-scale Simulations}


To evaluate the performance of \sys in a more realistic scenario, we run simulations with the web search workload~\cite{alizadeh2010data} and the data mining workload~\cite{greenberg2009vl2}. We use a three-tier fat-tree topology that consists of 8 core switches, 16 aggregator switches, 32 top-of--rack (ToR) switches, and 192 nodes. We create two fat-trees one with 10 Gbps links and the other with 40 Gbps links.
Network link delays are set 4us and host delays are set to 1 us. Maximum RTT between nodes is 52 us excluding queueing delay. To support multipath routing, Equal Cost Multi Path (ECMP) routing is used.
Flows start with interval of Poisson random variable. 
We set the parameters as recommended in the papers for DCTCP, HULL, and DX\@. For HULL, we set the target utilization to 0.95.

\begin{figure}[t]
\centering
\centering
\includegraphics[width=\columnwidth]{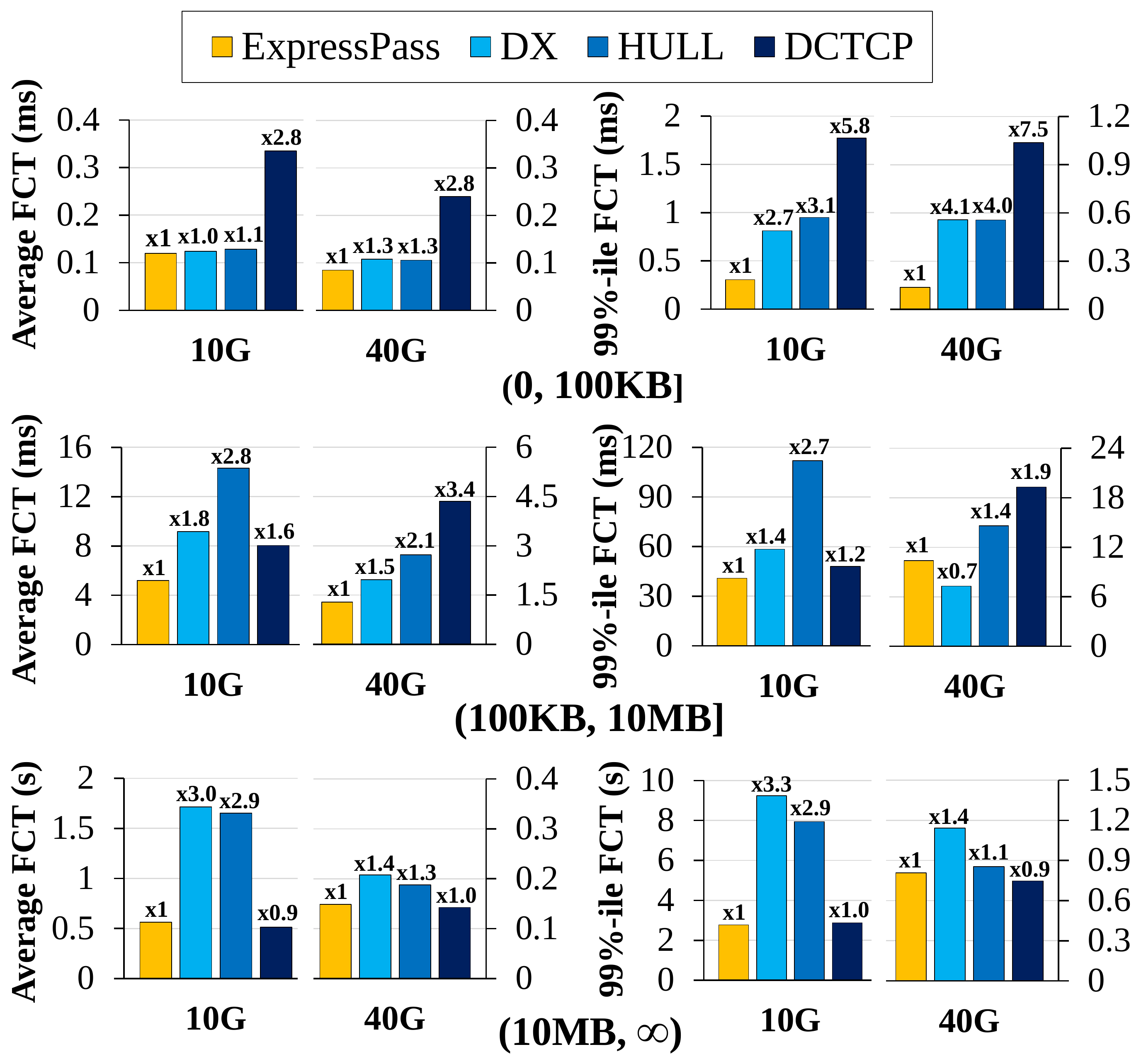}
\caption{FCT using data mining workload of 60\% load.}
\label{fig:macro_fct_10G}
\vspace{0.06in}
\end{figure}

Figure~\ref{fig:macro_fct_10G} shows the result for data mining workload in which 80\% of the flows are smaller than 100KB. For small flows under 100KB, \sys shows shortest FCT in both average and 99\%-ile. DX and HULL show similar FCT as \sys on the average, but their 99\%-ile FCT is two to four times slower. This is because \sys significantly reduces the FCT for small flows. The web search results (not shown in figure) show similar trends. For medium size flows between 100KB and 10MB, \sys outperforms all others except for DX at $40$~Gbps. For flows larger than 10MB, \sys is slightly slower than DCTCP by up to 10\%, while outperforming DX and HULL\@. 
As link speed increase to 40Gbps, \sys performs better. The flow FCT improvement is higher than that of the others. This is because \sys's convergence time is the same regardless of link speed, while others take longer to ramp up. Even the long flow's FCT is comparable to DCTCP\@.  
Overall, the simulation results clearly show the benefit of fast convergence for short flows. Our credit based mechanism effectively removes queueing while keeping the utilization high. 

We show the average and maximum queue length observed during the simulation in Table~\ref{tab:qlen_loss} for the data mining workload. On average, \sys uses less than one 10th of queueing compared to DCTCP, and uses similar amount as HULL and DX. Maximum queuing occupancy is lowest in \sys. 
All mechanisms did not incur any loss in the data mining workload. In contrast, in the web search workload, all but \sys exhibited packet losses. 


\begin{table}[t]
\centering
\resizebox{0.9\columnwidth}{!}{
\begin{tabular}{ccccc}
\toprule
  & \textbf{DCTCP} & \textbf{HULL} & \textbf{DX} & \textbf{ExpressPass} \\ \midrule
  Avg. (KB) & 10.08 & 0.826 & 0.851 & 0.938 \\ 
   Max (KB) & 243.1 & 40.0 & 307.6 & 29.9 \\
   \bottomrule

\end{tabular}
}
\caption{Queue occupancy 
per switch port in data mining workload (load 0.6) with 10Gbps link}
 \label{tab:qlen_loss}
\end{table}



\section{Discussion}
\label{sec:discuss}

\noindent
\textbf{Internal fragmentation of credit.} Because a sender can transmit up to an MTU size data packet per credit, if it sends less, difference between maximum and actual size of the packet will be wasted. In data mining (web search) workload, 727 (745) bytes per flow were wasted on average. Considering that the average flow size is 7.41 MB (1.65 MB), the size of internal fragmentation is only 0.01\% (0.04\%) of the flow.

\vspace{0.05in}
\noindent
\textbf{Ignored credit packets at the sender.} Currently, we let receivers  sending credit packets until the last data arrives assuming they do not know when the flow ends in advance. 
In data mining (web search) workload with 10Gbps link, each flow was wasting 30 (40) credits on average. Considering average flow size, the portion of the wasted credit packets is only 0.6\% (3.7\%). 
We believe this can be reduced if the end of the flow can be reliably estimated in advance. 

\vspace{0.05in}
\noindent
\textbf{Presence of other traffic.}
We assumed that we can control all traffic with \sys. Compatability with TCP is not our goal. However, in real datacenter networks,  some traffic, such as ARP packets, link layer control messages, may not be able to send credits in advance. One solution to this apply ``reactive'' control to account for the small amount of such traffic. When traffic is sent without credit, we absorb them in the network using queue and send credit packets from the receiver. This will drain the queue.

\section{Related Work.}
Our design is in part inspired by credit-based flow control~\cite{kung1994credit}, decongestion control~\cite{raghavan2006decongestion}, and pFabric~\cite{alizadeh2013pfabric}. Credit-based flow control is popular on high speed networks, such as  on-chip networks and Infiniband. Decongestion control and pFabric pioneered a design where hosts transmit data aggressively and switches allocate bandwidth. The difference is that we allocate bandwidth using credit. Finally, TVA~\cite{tva} use similar idea to rate limit requests at the router, but it is designed for DDoS prevention considering the size of response rather than congestion control.

DCQCN~\cite{zhu2015congestion} and TIMELY~\cite{mittal2015timely} are designed for datacenters that have RDMA traffic. DCQCN uses ECN as congestion signal and  QCN-like rate control. The main goals of DCQCN alleviate the problems caused by PFC by reducing its use while reducing ramp up time. 
TIMELY uses delay as feedback, similar to DX~\cite{lee2015accurate}, but  incorporates PFC to achieve zero loss and lower the 99\%-ile latency. 
PERC~\cite{jose2015high} proposes a proactive approach to  overcome the problems of reactive congestion control. We believe  \sys presents an alternative approach and shows promise in high-speed environment (e.g., 100 Gbps networks). 
 
\noindent\textbf{Flow scheduling in datacenter:} A large body of work focuses on flow scheduling~\cite{alizadeh2013pfabric,bai2015information,gao2015phost} in datacenter networks to minimize the flow completion times. Although the goals might overlap, flow scheduling is largely orthogonal to congestion control.  We believe congestion control is a more fundamental mechanism for network resource allocation. 
We also note that some flow scheduling schemes~\cite{munir2014friends} have been used in conjunction with congestion control to minimize the flow completion times. pFabric treats the network as a single switch and performs shortest job first scheduling assuming the flow size is known in advance. This requires switch modifications. PIAS~\cite{bai2015information} makes the approach more practical by  approximating shortest-job-first by implementing a multi-level feedback queue using priority queuing in commodity switches and does not require the knowledge of individual flow size. pHost\cite{gao2015phost} shares the idea of using credit (token) packets, but token packets in pHost is used for scheduling packets/flows rather than congestion control. It assumes a congestion-free network by using a network with full bisection bandwidth and packet spraying. In addition, pHost does require the knowledge of individual flow size in advance.

\noindent\textbf{Explicit congestion control:} Finally, some congestion control algorithms require in-network support~\cite{rcp,xcp,fcp}. These mechanisms introduce a form of in-network feedback with which switches explicit participate in rate allocation of each flow. To reach fast convergence, PERC~\cite{jose2015high} and  FCP~\cite{fcp} employ mechanisms for end-hosts to signal their bandwidth demand to the switches in the network, which require changes in the switches. In \sys, we use credit packets to signal demand and merely use rate-limiting which does not require modifications the switches.






\section{Conclusion and Future Work}\label{sec:conclusion}
In this work, we introduce \sys, an end-to-end credit-based congestion control. We extend the credit-based flow control 
to perform end-to-end congestion control in datacenters. We show that use of credit packets allows quick ramp-up of flows without worrying about data loss. By shaping the flow of credit packets in the network, \sys effectively controls congestion even \emph{before} sending data packets. By achieving fast convergence, it drastically reduces the FCT for small flows.  \sys requires a small amount of buffer to achieve high link utilization.  Our preliminary evaluation shows that \sys (1) outperforms other congestion control algorithms; (2) ensures high utilization;
and 3) takes better advantage of higher link speed than other mechanisms.
As future work, we plan to compare \sys with DCQCN, TIMELY, and PERC\@. 

\bibliographystyle{abbrv} 
\vspace{0.08in}
\bibliography{references}
\label{last-page}

\end{document}